\documentstyle[12pt,aaspp4,psfig,natbib]{article}

\newcommand{\Ly}{Ly$\alpha$}

\def\gtorder{\mathrel{\raise.3ex\hbox{$>$}\mkern-14mu
             \lower0.6ex\hbox{$\sim$}}}
\def\ltorder{\mathrel{\raise.3ex\hbox{$<$}\mkern-14mu
             \lower0.6ex\hbox{$\sim$}}}
\def\proptwid{\mathrel{\raise.3ex\hbox{$\propto$}\mkern-14mu
             \lower0.6ex\hbox{$\sim$}}}
\def\0946{PG~0946+301}

\def\arcsec{\ifmmode '' \else $''$\fi}

\def\arcsecpoint{\ifmmode ''\!. \else $''\!.$\fi}

\def\kms{\ifmmode {\rm km\ s}^{-1} \else km s$^{-1}$\fi}
\def\Msun{\ifmmode {\rm M}_{\odot} \else M$_{\odot}$\fi}
\def\Lsun{\ifmmode {\rm L}_{\odot} \else L$_{\odot}$\fi}
\def\Zsun{\ifmmode {\rm Z}_{\odot} \else Z$_{\odot}$\fi}

\def\ergscm2{ergs\,s$^{-1}$\,cm$^{-2}$}
\def\icm3{{\rm cm}^{-3}}
\def\icm2{{\rm cm}^{-2}}
\def\qo{\ifmmode q_{\rm o} \else $q_{\rm o}$\fi}
\def\Ho{\ifmmode H_{\rm o} \else $H_{\rm o}$\fi}
\def\ho{\ifmmode h_{\rm o} \else $h_{\rm o}$\fi}
\def\ltsim{\raisebox{-.5ex}{$\;\stackrel{<}{\sim}\;$}}

\def\vFWHM{\ifmmode v_{\mbox{\tiny FWHM}} \else
            $v_{\mbox{\tiny FWHM}}$\fi}
\def\CCF{\ifmmode F_{\it CCF} \else $F_{\it CCF}$\fi}
\def\ACF{\ifmmode F_{\it ACF} \else $F_{\it ACF}$\fi}
\def\Halpha{\ifmmode {\rm H}\alpha \else H$\alpha$\fi}
\def\Hbeta{\ifmmode {\rm H}\beta \else H$\beta$\fi}
\def\Hgamma{\ifmmode {\rm H}\gamma \else H$\gamma$\fi}
\def\Hdelta{\ifmmode {\rm H}\delta \else H$\delta$\fi}
\def\Lya{\ifmmode {\rm Ly}\alpha \else Ly$\alpha$\fi}
\def\Lyb{\ifmmode {\rm Ly}\beta \else Ly$\beta$\fi}
\def\Lyg{\ifmmode {\rm Ly}\beta \else Ly$\gamma$\fi}
\def\hi{H\,{\sc i}}

\def\ciii{\ifmmode {\rm C}\,{\sc iii} \else C\,{\sc iii}\fi}
\def\civ{\ifmmode {\rm C}\,{\sc iv} \else C\,{\sc iv}\fi}

\def\nv{N\,{\sc v}}

\def\o5007{[O\,{\sc iii}]\,$\lambda5007$}

\def\ovi{O\,{\sc vi}}

\def\o{\o}
\begin{document}

\title{ON THE COLUMN DENSITY OF AGN OUTFLOWS: \\
THE CASE OF NGC 5548 }


\author{
Nahum Arav\altaffilmark{1,2}, 
Kirk T. Korista\altaffilmark{3},
Martijn de~Kool\altaffilmark{4} 
}

\altaffiltext{1}{Astronomy Department, UC Berkeley, Berkeley, 
CA 94720, I:arav@astron.Berkeley.EDU}
\altaffiltext{2}{Physics Department, University of California, Davis, CA 95616}
\altaffiltext{3}{Western Michigan Univ., Dept. of Physics, 
Kalamazoo, MI 49008-5252}
\altaffiltext{4}{Research School of Astronomy and Astrophysics, ANU ACT,
 Australia}

\begin{abstract}

We re-analyze the {\em HST} high resolution spectroscopic data of the
intrinsic absorber in NGC~5548 and find that the \civ\ absorption
column density is at least four times larger than previously
determined.  This increase arises from accounting for the
kinematical nature of the absorber and from our conclusion that the
outflow does not cover the narrow emission line region in this object.
The improved column density determination begins to bridge the gap
between the high column densities measured in the X-ray and the low
ones previously inferred from the UV lines.  Combined with our findings
for outflows in high luminosity quasars these results suggest that
traditional techniques for measuring column densities: equivalent
width, curve-of-growth and Gaussian modeling, are of limited value
when applied to UV absorption associated with AGN outflows.

\end{abstract}

\section{INTRODUCTION}

Seyfert galaxies often show absorption features associated with
material outflowing from the vicinity of their active nuclei at
velocities of several hundred \kms\ (Crenshaw et~al.\ 1999).  These
features are traditionally observed in UV resonance lines (e.g.,
\civ~$\lambda\lambda$1548.20,1550.77,
\nv~$\lambda\lambda$1242.80,1238.82, \Ly), but with the improved X-ray
spectral capabilities of the {\em Newton} and {\em Chandra}
satellites, they are also detected in X-ray lines (Kaastra et~al.\
2000; Kaspi et al. 2000). Reliable measurement of the absorption
column densities in these lines are crucial for determining the
ionization equilibrium and abundances of the outflows, and the
relationship between the UV and the warm X-ray absorbers.

 NGC~5548 is one of the most studied Seyfert
galaxies, including intensive reverberation campaigns (Netzer \& Maoz
1990; Peterson et~al.\ 1991; Korista et~al.\ 1995), line studies (Goad
\& Koratkar 1998; Korista \& Goad 2000), and theoretical modeling (Done
\& Krolik 1996; Bottorff, Korista \& Shlosman 2000; Srianand 2000). The
intrinsic absorber in NGC~5548 was studied in the UV using the
{\em International Ultraviolet Explorer} (Shull \& Sachs 1993), the {\em HST}
Goddard High Resolution Spectrograph (GHRS) (Mathur, Elvis \& Wilkes
1999) and Space Telescope Imaging Spectrograph (STIS) (Crenshaw \&
Kraemer 1999), the {\em Far Ultraviolet Spectroscopic Explorer ( FUSE)}
(Brotherton et~al.\ 2001), and in X-ray with the {\em ASCA} (George et
al. 1998) and {\em Chandra} (Kaastra et~al.\ 2000) Satellites.  These high
quality observations combined with the relative simplicity of its
intrinsic absorption features make NGC~5548 a prime target for
understanding the nature of Seyfert outflows.

Our work on quasar outflows, has led us to suspect that the current
determination of column densities in Seyfert outflows is highly
uncertain.  In the last few years our group (Arav 1997; Arav et
al. 1999a; Arav et~al.\ 1999b; de Kool et~al.\ 2001; Arav et~al.\ 2001)
and others (Barlow 1997, Telfer et~al.\ 1998, Churchill et~al.\ 1999,
Ganguly et~al.\ 1999) have shown that in quasar outflows most lines are
saturated even when not black. We have also shown that in many cases
the shapes of the troughs are almost entirely due to changes in the
line of sight covering as a function of velocity, rather than to
differences in optical depth (Arav et~al.\ 1999b; de~Kool et~al.\
2001; Arav et~al.\ 2001a; we elaborate on this point below). As a
consequence, the column densities inferred from the depths of the
troughs are only lower limits. Furthermore, it is virtually impossible
to decouple the effects of optical depth from those of covering factor
for a single line, thus column density determinations in such cases
are not possible.  

In order to assess the importance of these effects in Seyfert
outflows, we reanalyze {\em HST} high-resolution spectra of NGC~5548.
Two such data sets exist  in the {\em HST} archive.  1) GHRS
observations including a 14,035 seconds exposure of the \civ\
intrinsic absorber obtained on 1996 August 24  (spectral resolution of
$\approx20,000$). These data are further described in Mathur, Elvis \&
Wilkes (1999). 2) STIS echelle spectrum using the E140M grating (4750
seconds exposure), which covers the \civ, \nv\ and \Ly\ intrinsic
absorption features, obtained on 1998 March 11 (full details are found
in Crenshaw \& Kraemer 1999).

\section{ANALYSIS}

\subsection{Methodology}

In studies of ISM and IGM absorption features we implicitly assume
that along our line of sight the absorber covers the emission source
completely and uniformly. Usually, in these environments this is an
excellent assumption since the geometrical size of the absorption
clouds are much larger than the emission sources they cover.  This
assumption allows us to directly convert the depth of the absorption
features to optical depth using $\tau=-\ln(I_r)$ (where $I_r$ is the
normalized residual intensity in the absorption trough), and from the
optical depth to obtain the column density. 

What happens when the complete and uniform covering assumption breaks
down?  For example let us assume that the absorbing material covers
only half of the emission source.  In this case even if the optical
depth of the absorber is very large (10, 100, 1000...) we would still
 detect an $I_r=0.5$.  But this $I_r$ would not give us any
information about the optical depth.  All it can tell us is that half
of the light-source's area is covered.  Furthermore, this geometrical
covering fraction can be velocity dependent.  In general, for cases
where the  line of sight covering of the light-source is not
complete, the depth of the trough at any given point will be a
convolution of the real optical depth and the covering fraction at
that velocity.

To account for partial covering we rely upon doublet lines that yield
two related sets of data from which the optical depth and
line-of-sight covering fraction can be solved for simultaneously
(Barlow \& Sargent 1997; Hamann et~al.\ 1997; Arav et al 1999b).  For
such analysis to succeed we must use unblended absorption features
seen in both components of the doublet, where the features are well
resolved spectroscopically and possess high enough signal to noise
ratio (s/n).  To make full use of these advantages we employ a simple
robust physical criterion. At each resolution element we require that
the real optical depth of the blue component is exactly twice that of the
red component, as dictated by the ratio of their oscillator strengths
(as is the case for the  \civ~$\lambda\lambda$1548.20,1550.77 doublet).
This is done after normalizing the data and putting the data set for
the two transitions on the same velocity scale.  The
absorption equations are:
\begin{eqnarray}
I_1(v)-(1-C(v)) & = & C(v)e^{-\tau(v)}  \\ I_2(v)-(1-C(v)) & = & C(v)e^{-2\tau(v)}, 
\end{eqnarray}
where $v$ is the velocity of the outflow, $I_1(v)$ and $I_2(v)$ are
the normalized intensities of the red and blue doublet components,
respectively, $C(v)$ is the effective covering fraction (see Arav et
al. 1999b), and $\tau(v)$ is the optical depth of the red doublet
component.  From these two equations we obtain:
\begin{equation}
C(v) = {{I_1(v)^2 - 2I_1(v) + 1}\over{I_2(v) - 2I_1(v) +1}} , \ \ 
\tau(v)=-\ln\left({I_1(v)-[1-C(v)]}\over{C(v)}\right)=-\ln\left({I_1(v)-I_2(v)}\over{1-I_1(v)}\right),
\label{eq:doublet}
\end{equation}
where the right most expression for $\tau(v)$ has not appeared in the
literature previously. Unlike its predecessor ($\tau(v)=f[C(v)]$), the new expression 
has a clear physical meaning and  behavior, with respect to the data. 

We note that this type of analysis can be employed to determine the
optical depth of ISM and IGM absorbers as well.  In these cases we
simply obtain $C(v)\approx1$ and $\tau(v)\approx-\ln[I_1(v)]$.

\begin{figure}
\centerline{\psfig{file=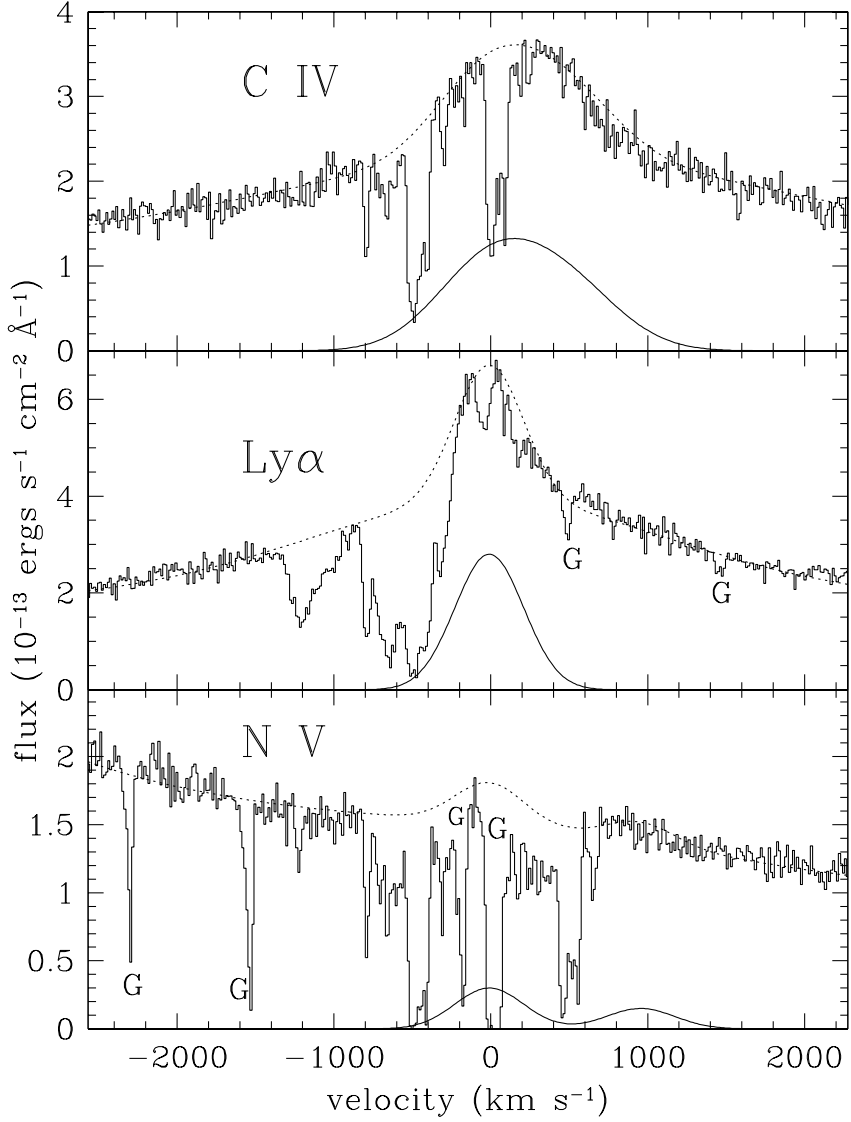,height=16.0cm,width=12.0cm}}
\caption{{\em HST}/STIS data of the NGC 5548 outflow  as seen in \civ\,
\Ly\ and \nv.  The data are boxcar smoothed over 5 pixels.  On top of
the data we plot our model for the unabsorbed emission.  This model
contains three contributions, a constant continuum level under the
emission line, a broad emission line (BEL) and a narrow emission line
(NEL) component.  The continuum and BELs are tightly constrained by
the data.  For the NEL we chose simple one-Gaussian models with the
least possible emission excess beyond what the data require, similar
in concept to the models chosen by Crenshaw \& Kraemer (1999).  The
NEL models are shown separately at the bottom of the figure. Galactic
absorption features are marked by ``G.''
}\label{fig1}
\end{figure}

\subsection{What Emission Components Does the Outflow Cover?}

In the spectral vicinity of the intrinsic absorber we find emission
from three different sources (see Fig. 1).  A continuum emission which
underlines the entire AGN spectrum,  Broad Emission Lines (BELs) with
a full width at half maximum (FWHM) of $\sim6000$ \kms\ and narrow
emission lines (NELs) with a FWHM of up to 1000 \kms (for
\civ~$\lambda$1549).  Before attempting to extract the column density
from the absorption troughs it is essential to establish which
emission components the intrinsic absorber covers.  If we assume that
the absorber covers all the emission sources when in fact it only
covers the continuum and the BEL, we might greatly underestimate the
optical depth and hence the column density in the absorption troughs.

To assess which emission components the outflow is covering, we
constructed models for the unabsorbed emission for the \civ, \Ly\ and
\nv\ lines (see Fig. 1).  Potentially, there is a lot of freedom in
such emission models.  The centroid redshift of the emission line
components can vary between lines and between the  NEL and BEL
components of the same line, and the considerable amount of absorption
makes the NEL model quite subjective (see Mathur, Elvis \& Wilkes
1999).  In order to minimize the uncertainty we use a simple and
restrictive approach while constructing the emission components.  We
use only Gaussian components, which are all centered on the systemic
redshift of the object (z $=$ 0.01717 from the NASA/IPAC Extragalactic
Database [NED], based on \hi\ observations). To obtain a physically
realistic model, each doublet component in \civ\ and \nv\ is modeled
by separate Gaussian components.

The continuum and BELs are tightly constrained by the data.  We obtain
a very good agreement with the \civ\ and \Ly\ data by modeling both
BELs with two Gaussian components characterized by FWHM of 2500 and
8000 \kms. Two Gaussian components are the minimum number necessary to
fit the data. This modeling also agrees well with Brotherton et
al. (1994), who argued that most AGN \civ\ BEL profiles are well fit
by two Gaussian components with FWHM of 2000 \kms\ and 8000 \kms.  For
the NELs we chose single-Gaussian models for each component of the
doublet, with the least possible emission excess beyond what the data
require, similar in concept to the models chosen by Crenshaw \&
Kraemer (1999).  These models are constructed so no substantial
emission features lie above the model, which is a necessary condition
for the model to be valid.  There are two reasons for choosing a NEL
model with the least possible emission excess beyond what the data
require.  First, it is always possible to choose an emission model
that is substantially higher than the data and attribute the
difference to absorption.  This can be done for all the emission
components and for any arbitrary wavelength range.  However, the
greater the excess emission we invoke the less plausible our model
becomes since we have to invoke an increasing amount of unnecessary
absorption.  Second, as we show below, such models put the most
stringent test on the question of whether or not the outflow covers
the NEL.

The best fit is obtained with a
660 \kms\ FWHM for each \civ\ component and 430 \kms\ for \Ly.  For
\nv, the unabsorbed emission model is dominated by the red wing of the
\Ly\ BEL.  This leaves little room for a wide BEL.  Instead we used a
small BEL component with a FWHM of 2500 \kms\ coupled with the maximum
allowable NEL with a 430 \kms\ FWHM (which is constrained quite tightly
by the fit of its red component to the absorption-free data around
--1000 \kms, see Fig.~1). We note that coincidently, Brotherton et
al. (2001) modeled the \ovi\ NEL in their {\em FUSE} data as either a ``wide''
NEL with FWHM of 658 \kms\ or a ``narrow'' NEL with FWHM of 432 \kms.
From the depth of the strongest absorption features associated with
the outflow, we deduce that the outflow must fully cover the BEL and
continuum regions. The important issue to resolve is whether the
outflow also covers the NEL region.
 
A decisive way to establish whether the NEL region is covered by the
outflow is to find a black (i.e., zero flux) absorption feature on top
of the NEL. By inspection of figure 1, this is not the case in
NGC~5548.  Lacking this definitive test we turn to the next strongest
test.  By choosing NEL models with the least possible emission excess
beyond what the data require, any significant deepening of the troughs
below these models necessitates significant coverage of the NEL region
by the outflow. From figure 1 we infer that, within the level of
noise, in all three lines the troughs do not deepen below the NEL
model. A similar situation exists in the {\em FUSE} data of the
\ovi~$\lambda\lambda$1032,1038 in this object (Brotherton
et~al. 2001).  Although these occurrences are not sufficient to prove
the no-NEL-coverage picture, they are strongly suggestive that this is
indeed the case. If the NEL region is covered by the flow, it is
highly probable that the absorption in at least one of these four lines
would be deep enough to categorically show that the NEL must be
covered.  Furthermore, we point out that in all three lines shown in
figure 1, the deepest absorption is very close to the inferred NEL
flux level.  This is quite a remarkable coincidence if the NEL region
is covered.  However, this coincidence becomes a natural consequence
of a picture where an optically thick outflow fully covers the
continuum and the BEL regions, but very little of the NEL region. On
the basis of these two arguments we conclude that it is most likely
that the outflow covers the continuum and the BEL regions but not the
NEL region. In \S~2.4 we present further observational support for
this picture.

Such a scenario is also plausible on physical grounds.  Reverberation
studies of NGC~5548 indicate a continuum source whose radius is
$\ltsim$1 light day, and a \civ\ BEL region whose luminosity-weighted
radius is $\sim$30 light days (Clavel et~al.\ 1991; Krolik et~al.\
1991; Korista et~al.\ 1995; Kaspi \& Netzer 1999; Korista \& Goad
2000). In contrast, the NEL region is larger than $\sim$10 light years
across (Wilson et~al.\ 1989). The very large difference in sizes
comfortably allows for a model where the outflow covers most or all of
the relatively small continuum source and BEL region, while not
covering the hundred times larger NEL region.  BeppoSAX observations
of NGC~5548 (Nicastro et~al.\ 2000) suggest that the size of the warm
absorber is $\sim$20 light days.  Since it is likely that the UV and
X-ray absorbers are connected, these findings lend additional support
for the no-NEL-coverage picture.

\begin{figure}
\centerline{\psfig{file=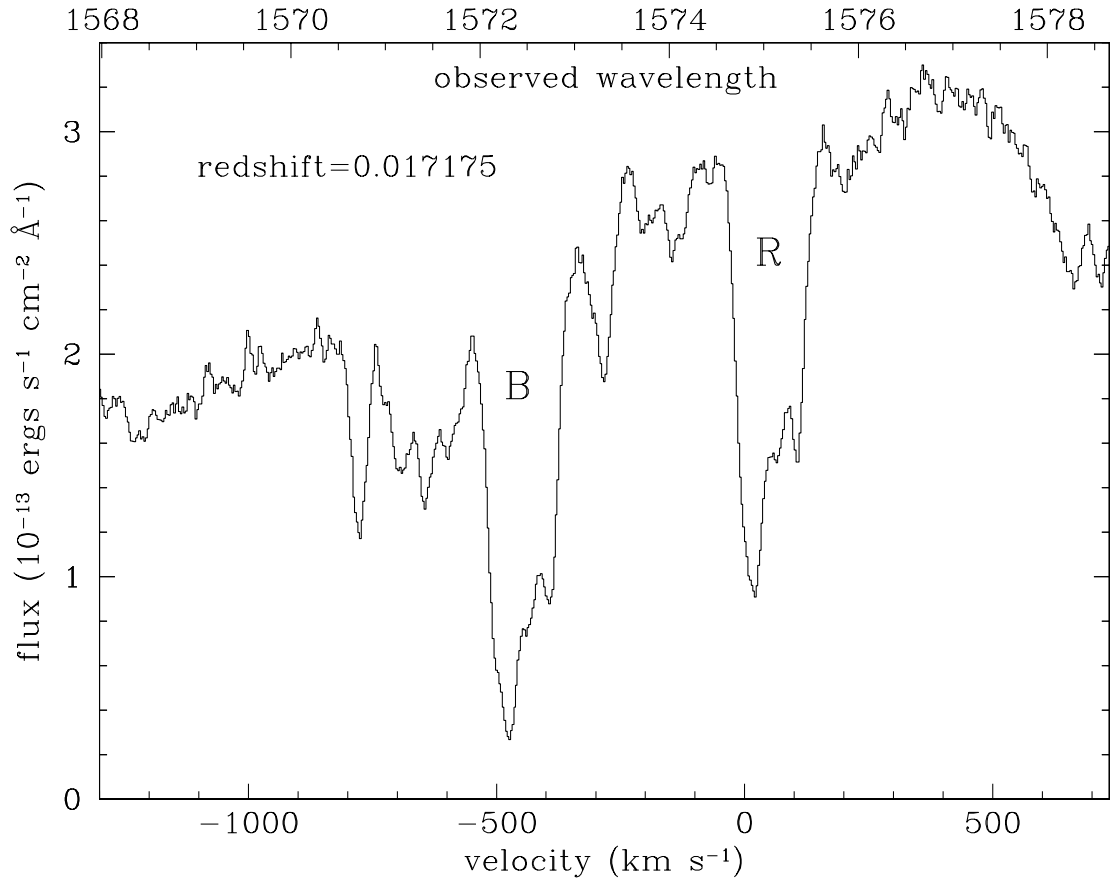,angle=-90,height=12.0cm,width=16.0cm}}
\caption{{\em HST}/GHRS spectrum of the \civ~$\lambda$1549 region in NGC
5548.  The data are boxcar-smoothed over 5 pixels.  Our analysis in
this letter concentrates on the deepest absorption component of the
outflowing material, seen in both components of the \civ\ doublet
(marked with B and R for the blue and red components, respectively).
}\label{fig2}
\end{figure}

\begin{figure}
\centerline{\psfig{file=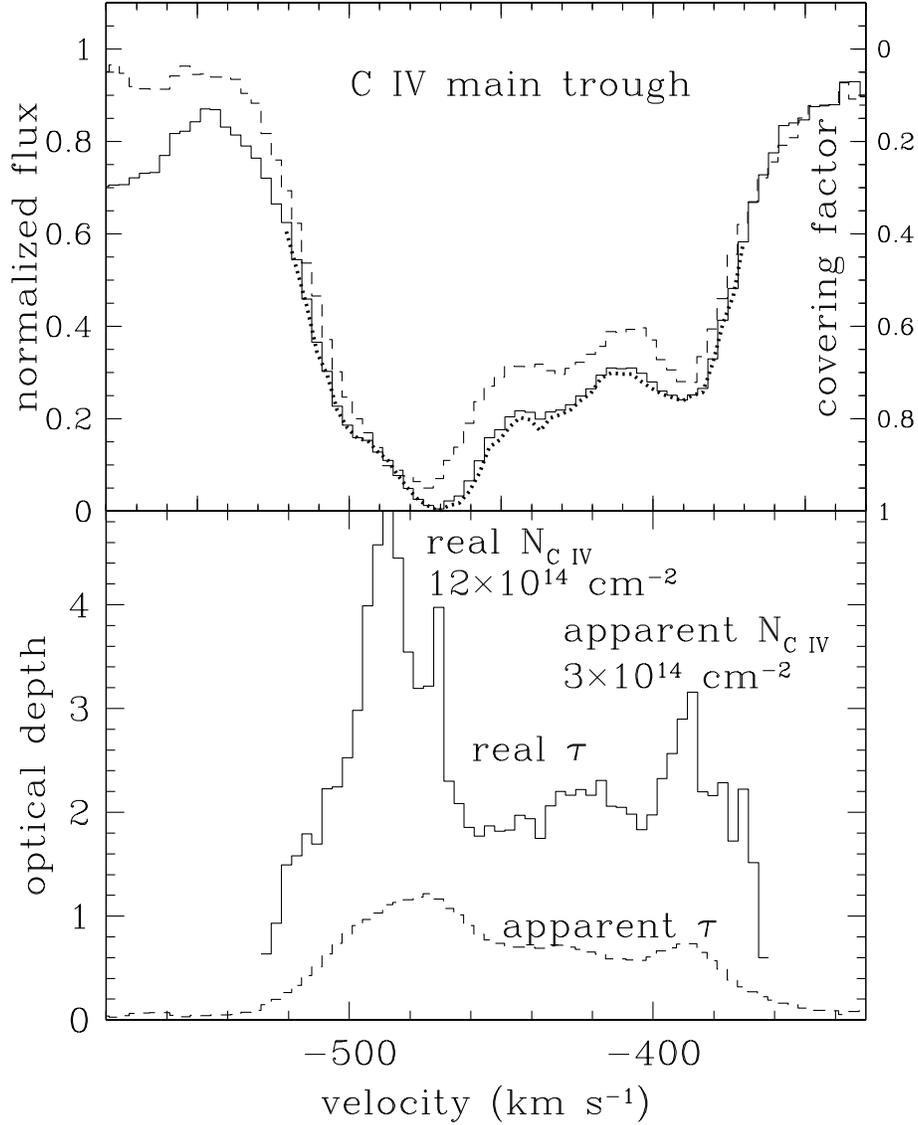,height=16.0cm,width=12.0cm}}
\caption{Solutions for the covering factor and real optical depth for
the case where the narrow emission line is not covered by the
outflow. The histograms in the upper panel show the normalized data,
for the deepest subtrough in the \civ\ absorption complex (B and R in
Fig~\ref{fig2}, boxcar-smoothed over 5 pixels),  where the solid and dashed histogram
corresponds to the blue and red  doublet components.  
  The dotted line is the covering factor
solution, which demonstrates that the shape of the trough is
determined almost entirely by the covering factor, which is a strong
function of velocity.  In the lower panel we show the real optical
depth, with the apparent one for comparison.  The integrated \civ\
column density is four times larger than the apparent one. 
}\label{fig3}
\end{figure}

\subsection{Optical Depth and Covering Fraction Solutions}

Due to their wide spectral coverage, the STIS data discussed above are
valuable for trying to determine which emission component is covered
by the outflow. However, the s/n  of these data is not high
enough to obtain clear results with the analysis method described in
\S~2.1.  Luckily, there is another set of high-resolution spectral data
of the intrinsic absorber in NGC~5548, with sufficient s/n  for
the $\tau(v)/C(v)$ analysis.  These are {\em HST}/GHRS data, which show the
outflow only in \civ~$\lambda\lambda$1548.20,1550.77 (Mathur, Elvis \&
Wilkes 1999; see Fig~\ref{fig2}). We concentrate our analysis on the
deepest subcomponent of the outflow, which, due to its depth, is less
sensitive to the emission model's uncertainties.  This is component
number 4 in the Crenshaw \& Kraemer (1999) designation.  We mark the
blue and red subtroughs of this components with ``B'' and ``R'' in
figure \ref{fig2}.

In figure \ref{fig3} we show the results of the analysis, which
assumes that the NEL is not covered by the flow.  The normalized data
for this case is given by: $I_r(\lambda)=(F-NEL)/(EM-NEL)$, where $F$
is the observed flux, $NEL$ is the NEL model and $EM$ is the full
non-absorbed emission model.  In order to derive the optical depth and
covering factor functions, we use the same velocity frame for the
``B'' and ``R'' subtroughs.  This is done by shifting the wavelength
of the ``R'' subtrough to
$\lambda_{\mbox{shift}}=\lambda\times1548.20/1550.77$ and transferring
both subtroughs to a velocity presentation using the systemic redshift
of the object. Figure \ref{fig3} shows that the residual intensities
in the doublet components do not adhere to the expected 1:2 optical
depth ratio (which implies $I_{\mbox{blue}}=I^2_{\mbox{red}}$), but
are much closer to a 1:1 ratio.  This has two effects: 1) To restore the 1:2
optical depth ratio the covering fraction has to be very close to the
$I_{\mbox{blue}}$ curve, which unequivocally demonstrates that the
shape of the trough is determined almost solely by the covering
fraction at each velocity. 2) Under these conditions, the optical
depth is much higher than the apparent optical depth defined as
$\tau_{ap}=-0.5\ln[I_{\mbox{blue}}]$ (see eq.~2), which assumes
complete coverage.  The comparison between the real and apparent
optical depth is shown in the bottom panel.  We note that the column
density determination for this component in both previous analyses of
the GHRS data ($2.9\times10^{14}$ cm$^{-2}$: Crenshaw \& Kraemer 1999;
$3.6\times10^{14}$ cm$^{-2}$: Mathur, Elvis \& Wilkes 1999) are very
similar to the one we extract from the apparent optical depth
($3\times10^{14}$ cm$^{-2}$).  However, the column density inferred
from the real optical depth solution is $12\times10^{14}$ cm$^{-2}$.
A similar $\tau(v)/C(v)$ analysis of the STIS data yields consistent
(albeit noisier) results. 

What are the uncertainties in the derived solutions?  There are two
sources of error in the $\tau(v)/C(v)$ analysis 1) data uncertainties
due to the finite S/N in each trough.  2) errors in the unabsorbed
emission model.  In principle, all three error sources (two for the
blue and red troughs data and one for the emission model) should be
propagated through equation array (\ref{eq:doublet}) in order to
derive quantitative error estimates.  In practice we find that the
systematic uncertainties in the NEL model dominate the total error.
The smoothed data we use have a minimum S/N=10 at the bottom of the
blue trough and generally a S/N$\gtorder20$ across most of the extent
of both troughs.  Since the relative difference between the normalized
intensities is mostly 20--30\% (see Fig.~\ref{fig3}), the errors due
to S/N are modest. This can also be inferred from the smooth behavior
of the $\tau$ solution in Fig. \ref{fig3}.  

Estimating the uncertainties in the unabsorbed emission model is much
more complicated.  Variation in the NEL model affect the red and blue
troughs in different ways and there is no physical reason to assume
that the NEL model must be Gaussian in nature.  We experimented with
many NEL models that adhere to the ``least possible emission excess''
principle and are Gaussian in nature.  In general these models gave
physical solutions under the no-NEL-coverage scenario i.e.,
$I_1,I_2>0$ and $I_1^2\leq I_2\leq I_1$.  We estimate that the NEL
model uncertainties induce a maximum error of 10\% in $\Delta I/I$,
over most of the trough.  This translates to a 20\% error in $\tau$
and therefore in our deduced column density.  However, in the region
of high $\tau$ (--500 to --470 \kms) the error can drive the $\tau$
solution to infinity (since it can make $I_2\geq I_1$).  We therefore
note that the column density might be significantly higher than the
value quoted above. Only significantly improved data will allow us to
clarify this issue.  For completion we mention that the errors
in $C(v)$ are very small due to the mathematical properties of the
$C(v)$ solution (see equation array (\ref{eq:doublet})). Whenever
$I_2$ is significantly larger than $I_1^2$ (but still obeys $I_2\leq
I_1$), $C(v)\approx1-I_2$.

We have also solved for the optical depth and covering factor for the
case when the NEL is assumed to be covered by the flow.  For this case
we find that the shape of the trough is still largely determined by
variation of the covering factor as a function of velocity, and that
the real optical depth is moderately higher than the apparent one. The
resultant difference in integrated column density is roughly 50\%. 

\subsection{Additional Support for the Moderate Saturation Picture}

\begin{figure}
\centerline{\psfig{file=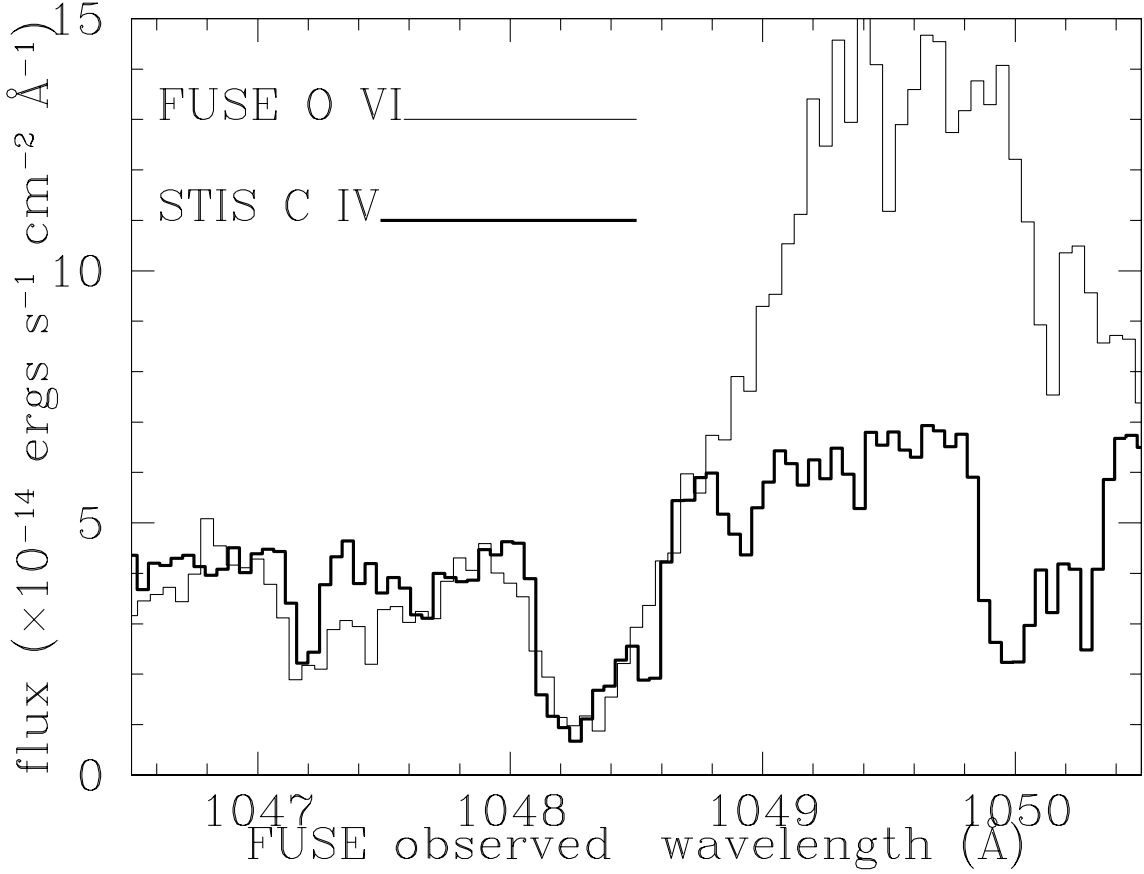,angle=-90,height=12.0cm,width=16.0cm}}
\caption{A Comparison between the STIS data of the \civ\ trough and
the {\em FUSE} data of the \ovi\ trough.  The blue component of the
\civ\ trough was shifted to the observed frame of the blue component
of the \ovi\ data.  The flux scaling of the \civ\ data is designed to
match the red wing of the \ovi\ flux level in order to obtain a direct
comparison which is not distorted by the very different NELs.  The
main trough (1048.0--1048.7 \AA) is almost identical in shape and depth
for both \civ\ and \ovi.  This is a powerful and simple indication
that both troughs are at least moderately saturated.}\label{fig4}
\end{figure}

Recent {\em FUSE} observations of NGC~5548 provide an independent
indication for saturation in the main absorption trough.  These data
were taken on June 7 2000 and are described by Brotherton
et~al.~(2001).  A direct comparison between the {\em FUSE} and STIS
observation depends on the assumption that the main trough did not
vary significantly between the {\em FUSE} and STIS observing epochs
(roughly two years).  In the case of NGC~5548, this assumption is
supported by noting that the main \civ\ trough shows very little
variation between the STIS and GHRS observing epochs, which also span
roughly two years (see \S~1).  

Assuming no significant variability, a comparison between the {\em
FUSE} observations and the STIS data strongly suggest that the main
\civ\ and \ovi\ troughs are saturated (see Fig.~ \ref{fig4}).  The
shape and depth for the main absorption feature is almost identical in
both \civ\ and \ovi.  Such a coincidence is highly unlikely if the
depths of the troughs are due to real optical depth effects.  There is
no a-priori reason to expect that the optical depth of lines that
arise from ions with very different ionization potential and from
elements with differing chemical abundances, should be almost
identical.  However, this situation is naturally explained if the
shape of the troughs are determined mainly by variations of the
covering fraction as a function of velocity.  In this case, the
optical depth of the \civ\ and \ovi\ can be quite different but the
resulting trough shapes can be very similar.  This independent
indication for saturation in the \civ\ trough supports our
no-NEL-coverage picture.  In order to have the covering factor
dominating the shape and depth of the \civ\ trough, we require the
residual intensities of the blue and the red doublet components to be
close numerically.  This is the case when the NEL is not cover (see
Fig.~\ref{fig3}), but is not so if the NEL is covered (as can be
inferred from Figs.~\ref{fig1} and \ref{fig2})

\section{DISCUSSION}

Previously published analyses of the GHRS and STIS data (Mathur, Elvis
\& Wilkes 1999; Crenshaw \& Kraemer 1999) modeled the absorption
features as Gaussians, using the software package SPECFIT (Kriss 1994).
The possibility of partial covering factor was acknowledged and
attempts were made to measure it. However, these analyses neglected two
important effects. First, the covering factor can be a strong function
of velocity, whereas these studies assumed that it is constant across
each absorption component. Second, there is the possibility that the
outflow covers the continuum and the BEL, but not the NEL. As our
analysis show these two effects are important and have a profound
influence on the deduced ionic column densities in the absorption
troughs.

The difference between the two approaches stems from a fundamental
conceptual issue which is ignored by the Gaussian fitting technique:
{\it each absorption feature is not a single cloud, it is an outflow
component}. This is because the absorption components are more than
100 \kms\ wide (even thousands in quasar outflows). These cannot be
single clouds since, for acceptable photoionization equilibria,
clouds' thermal widths are $\ltorder10$ \kms. It is reasonable to
expect that a cloud will have a single covering factor.  However, this
is not the situation for an outflow, where the covering factor can be
a strong function of velocity.  Indeed our analysis of the outflows in
QSO~1603+3002 (Arav et~al.\ 2001) and QSO~1044+3656 (de Kool et
al. 2001) show precisely this effect.  Toy models which illustrate
these kinematic/geometric effects are found in Arav (1996) and Arav et
al. (1999b).  A similar picture arises in global models for the
structure of AGN (Elvis 2000; Elvis 2001). Since the absorption
components are not single clouds, there is no physical basis for
modeling them as such in the form of Gaussians.

In the Seyfert galaxy NGC~5548, as well as for outflows in high
luminosity quasars, we have shown that the fundamental quantity of
column density can be greatly underestimated if one uses the
traditional measuring methods of equivalent width, curve-of-growth and
Gaussian modeling.  It is rather difficult to modify these techniques
to account for the effects we describe here.  The main obstacle is the
strong dependency of the covering fraction upon velocity. At this
point we do not foresee any way to adequately account for
velocity-dependent covering fraction in the traditional methods,
without fully solving for it as done here.

What are the implications of the results presented here on our
understanding of AGN outflows?  Almost all our deductions about the
outflows starts from the measured column densities.  For example, the
existing column density measurements in NGC~5548 coupled with
photoionization modeling, suggests that the nitrogen to carbon ratio
is significantly higher than solar (Srianand 2000).  However, if the
\civ\ column density is much higher than previously thought, the
nitrogen overabundance may not be needed.  Dynamical models of the
winds are also strongly affected since the mass and energy fluxes are
proportional to the column densities (Arav, Li \& Begelman 1994;
Proga, Stone \& Kallman 2000). One of the most important implications
is in regards to the connection between the UV and the warm X-ray
absorber.  Based on the existing UV column density measurements it
appears that in NGC~5548 the two absorbers cannot arise from the same
gas, since the X-ray column densities appear to be a hundred times
higher (Kaastra et~al.\ 2000) than those extracted in the UV.  (We note
that reanalysis of the {\em Chandra} data using improved calibration,
suggest that the equivalent width of the X-ray absorption lines are
roughly half of the initial measured values; Nicastro et~al.\ 2001).
However, if the UV column densities are strongly underestimated, this
situation might change.

Finally, we speculate that the difference of two orders of magnitude
between column densities inferred from UV and X-ray observation of
intrinsic absorbers might be primarily an illusion.  Almost all the
intrinsic absorption troughs seen in Seyferts are not black (see
Crenshaw et al. 1999).  Therefore, their apparent optical depth is
$\ltorder3$.  To be detectable in X-ray spectra, troughs must have
$\tau\gtorder1$ (see Kaastra et al. 2000). Usually it is assumed that
the depth of a given trough is indicative of its column density.
Therefore, for similar oscillator strength, a similarly shaped trough
(i.e., residual intensity as a function of velocity) yields roughly a
hundred times higher column density in an X-ray line compared to a UV
line simply because its wavelength is hundred times smaller (since
$N_{ion}\propto\lambda^{-1}$).  This remarkable situation will occur
every time we will detect the same trough in both UV and X-rays (e.g.,
Kaastra et al. 2000).  If however, the shape of the UV troughs has
little to do with real optical depth, then their column densities can
be much higher, averting this unlikely coincidence.

\clearpage

\section{SUMMARY}

We have reanalyzed the available {\em HST} high-resolution spectra of
the intrinsic absorber in NGC~5548; taking into account the kinematic 
origin of the troughs. For the highly probable scenario where the   
outflow does not cover the narrow emission line region we find that:

\begin{enumerate}

\item The \civ\ column density associated with the deepest trough
is at least four times larger than previously estimated.

\item The shape of this trough is almost entirely due to changes in
the line of sight covering as a function of velocity, rather than to
differences in optical depth, similar to our findings for troughs in
quasar outflows.

\item Our findings suggest that traditional techniques for measuring
column densities: equivalent width, curve-of-growth and Gaussian
modeling, are of limited value when applied to UV absorption associated
with AGN outflows.

\item We speculate that the large differences in column densities
inferred from UV and X-ray detection of the same intrinsic absorption trough,
may be primarily an  illusion, since the depth of the UV trough
is a poor indicator of its real optical depth.

\item In addition, for absorption doublets with 1:2 oscillator
strength ratio, we introduce a physically intuitive solution for
$\tau(v)$ under the partial covering formalism.

\end{enumerate}

\end{document}